\documentclass[final]{ias2}

\usepackage{graphicx} 
\usepackage{multirow}
\usepackage{array} 

\usepackage{hyperref} 

\begin{document}

\markboth{Avalanche transmission and critical behavior in load bearing hierarchical networks}{Ajay Deep Kachhvah, et. al.}

\title{Avalanche transmission and critical behavior in load bearing hierarchical networks}

\author[sin]{Ajay Deep Kachhvah} 
\email{ajay@physics.iitm.ac.in}
\author[sin]{Neelima Gupte} 
\email{gupte@physics.iitm.ac.in}
\address[sin]{Department of Physics, Indian Institute of Technology Madras, Chennai - 600036, India}

\begin{abstract}
The strength and stability properties of hierarchical load bearing networks and their strengthened variants have been discussed in recent work. Here, we study the avalanche time distributions on these load bearing networks. The avalanche time distributions of the $V-$ lattice, a unique realization of the networks, show power-law behavior when tested with certain fractions of its trunk weights. All other avalanche distributions show Gaussian peaked behavior. Thus the $V-$ lattice is the critical case of the network. We discuss the implications of this result.

\end{abstract}


 
\maketitle


\section{Introduction}

It is well known that many important systems such as the World Wide Web, and the Internet, power grids, cellular networks can be modeled as complex networks \cite{strogatz, barabasi}. Branching hierarchical networks 
constitute an important class of networks, and have been used as models of granular media \cite{copper}, river networks \cite{river}, as models of the lung inflation process \cite{inflation, inflation2}, directed percolation processes \cite{domany} and other \cite{voter}. Branching hierarchical networks 
have also been studied as models of load bearing networks, and strategies 
to improve their strength \cite{janaki} as well as their tolerance to failure \cite{ajay} have been studied in detail. In this paper, we study avalanche propagation on the load bearing branching hierarchical network, and show that the distribution on a specific realization of the lattice, known as the $V-$ lattice shows power-law behavior. Thus the $V-$ lattice constitutes the critical case of the load-bearing networks. We also discuss the relevance of this result in application contexts.

\section{Hierarchical models}

The distributions of avalanche time are studied for a $2-D$
hierarchical lattice, henceforth to be called the original lattice, two
of its variants enhanced by capacity enhancing strategies \cite{janaki},
and also for a specific realization of the original lattice, called the $V -$ lattice.

\begin {figure}
\begin{center}
\includegraphics[height=5cm,width=12cm]{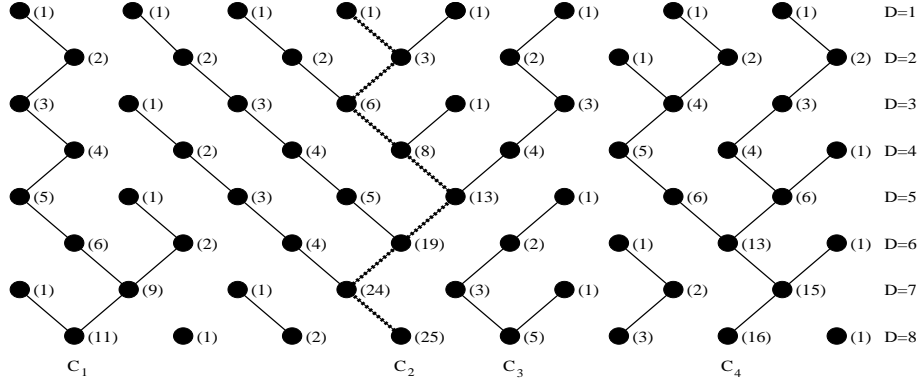}
\caption{\label{fig:lattice} A network of M=8 layers with 8 sites per layers with connection-probability $p=1/2$. The beaded line is the trunk of maximal cluster. The weight bearing capacity of the trunk is ${\it W_T}$=99.}
\end{center}
\end{figure}

The original lattice is, in fact, the special $q(0,1)$ case of the Coppersmith model \cite{copper} of granular media and Scheidegger's river model \cite{river}, in which, a site in a layer is connected randomly to one of its two neighbors in the layer below, yielding a hierarchical river like branched structure.
We discuss here the version of the lattice set up in Ref. \cite{janaki}.
Each site in the lattice has the capacity to support unit weight if it is not connected to any site in the layer above, and has capacity $w+1$ if it is connected to sites whose capacities add up to $w$, in the layer above. Thus, the capacity $w(i^M)$ of the $i^{th}$ site at any layer $M^{th}$ is given by
\begin{equation}
w(i^M)=l(i^{M-1}_l,i^M)w(i^{M-1}_l)+l(i^{M-1}_r,i^M)w(i^{M-1}_r)+1
\end{equation}
where ${\it i_l^{M-1}}$ and ${\it i^{M-1}_r}$ are the left and right neighbors of the site $i^M$, in the ${\it M-1}$th layer.
The quantity $l(i^{M-1}_l,i^M)$ takes the value $1$ if a connection exists between ${\it i^{M-1}_l}$ and ${\it i^{M}}$ and $0$ if a connection does not exist.
Fig.\ref{fig:lattice} shows a realization of a original network with
lattice side $M=8$. It is clear that the network consists of many
clusters, where a cluster consists of a set of sites connected with each
other. The largest such cluster of sites in the network is called the
maximal cluster of the network. Fig.\ref{fig:lattice} shows a
representation the network where the connections between sites are shown
by solid lines, and the numbers in the brackets indicate the capacity of
each site. $C_{1}, C_{2}, C_{3}$ and $C_{4}$ are the clusters seen in this realization of network, and $C_2$ is the maximal cluster. The beaded line denotes the trunk of maximum cluster, where the trunk is the set of connected sites with highest weight bearing capacity in the maximal cluster.
 
\subsection{The $V-$ Lattice Network}
\begin {figure*}
\begin{center}
\includegraphics[height=5cm,width=12cm]{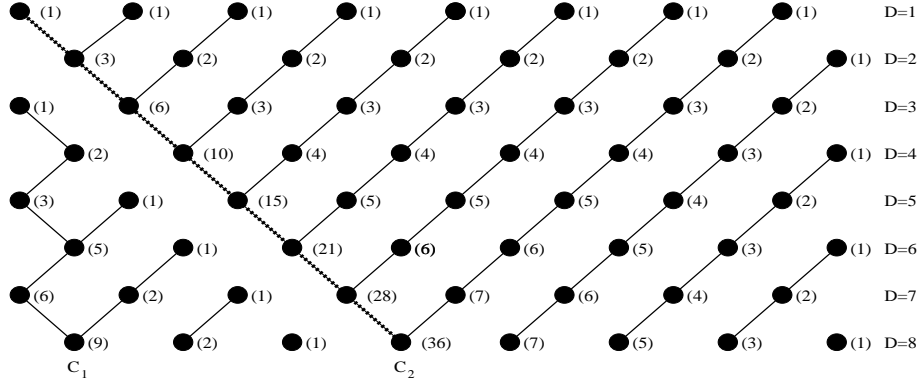}
\caption{\label{fig:vlattice} The $V-$ lattice network of $D$=8 layers with 8 sites per layers which is the critical case of our original network. The beaded line is the trunk of maximal cluster with weight bearing capacity ${\it W_T}$=120.}
\end{center}
\end{figure*}                        
The original lattice has a very special realization which bears
the maximum trunk capacity compared to all possible  realizations. This
lattice bears a unique $V-$ shaped cluster that includes all the sites
in the first layer, and $(M-I+1)$ sites in the $I-$th layer, where $M$
is the total number of layers. One of the arms of the $V$ constitute the
trunk, and all other connections run parallel to the arm of the $V$ that
is opposite to the trunk. Thus, this cluster includes the largest number
of sites, and is thus the largest possible cluster the original lattice
could have. We call this lattice the $V-$ lattice, and the cluster the
$V-$ cluster. The $V$-lattice realization with lattice side $M=8$ is shown in Fig. \ref{fig:vlattice}. The largest cluster here is the cluster $C_2$. Every site in the cluster at the layer $I$ has capacity $I$, except for the trunk site which has capacity $w_T(I)=w_T(I-1)+I$ in the $I$-th layer. We will see that this network corresponds to the critical case of the original lattice.

\section{Probability distributions of avalanche times}

It is interesting to study the avalanche phenomena on the network. Our study is of relevance to any situation where threshold phenomena propagate on a network. Examples of this range from power propagation on grids, electrical impulses on neural networks, ventilation in respiratory networks to directed percolation and granular media.

We define avalanches on the network in terms of weight transmission.
The weight transmission in the network takes place along the connections between  sites. When a site
in the first layer of the network receives a weight ${\cal W}$, 
it retains an amount equal to its capacity $W_c$ and transmits the rest, i.e. ${\cal{W}}-W_c$, to the site it is 
connected to, in the layer below. Thus, the weight transmission is in 
the downward direction and the sites involved in this process
with their connections constitute the path of transmission. Let $P$ 
be one such path and $P_{D}$ be the site on $P$ in the $D^{th}$ layer.
Then, the excess weight at a site $P_D$ in the $D^{th}$ layer is given by:
$$W^{ex}(P_D)= {\cal{W}}-\sum_{K=1}^{D}~W_c(P_K).$$ 
If $W^{ex}(P_D) \le 0$, then the transmission ends at the $D^{th}$ layer
of the path $P$ and is considered to be successful. On the other hand,
if $W^{ex}(P_D) > 0$, the weight is transferred to $P_{D+1}$. 
Finally, if there is still excess weight left at the $M^{th}$ layer, it
is then transmitted to the corresponding site in the first layer and the
second cycle of downward transmission begins as described above. This process continues till either there is no excess weight left,i.e. the transmission is
successful, or the receiving site is not able to transmit the excess to
the site in the layer below, and the transmission fails.  
Such a failure occurs when the transmitting site is
connected to a site that has already received its share of the
weight (i.e. saturated it's capacity) in the first cycle of transmission,
thus making further transmissions impossible. This process of
weight transmission  is defined as an avalanche. The time
taken for an avalanche is defined to be the number of layers traversed by
the weight in the network.

In this paper we compare the probability distribution of avalanche times $t$ between the original lattice, and its strongest realization, the $V-$ lattice. This probability distribution as the name suggests is, in fact, the distribution of the number of layers traversed during all cycles of successful avalanche transmission by a test weight placed at a random site in the first layer for any lattice. This probability distribution in case of the original lattice tested with the trunk weights, has been studied by Janaki $et.$ $al$ \cite{janaki}.
\begin{figure}[ht]
\begin{center}
\includegraphics[width=0.6\columnwidth]{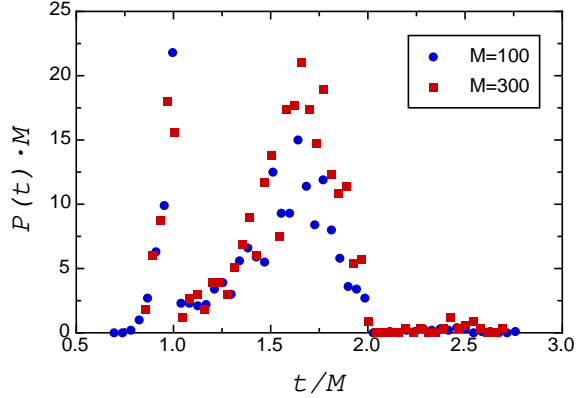}
\caption{Scaled probability distribution of avalanche time $t$ for original lattice corresponding to $1000$ realizations of networks of side $M=100$, and $M=300$. Both, the distributions for $M=100$ and $M=300$ scale by the total number of layers $M$ in network.}
\label{fig:orig}
\end{center}
\end{figure}

\begin{figure}[ht]
\begin{center}
\begin{tabular}{cc}
\includegraphics[width=0.45\columnwidth]{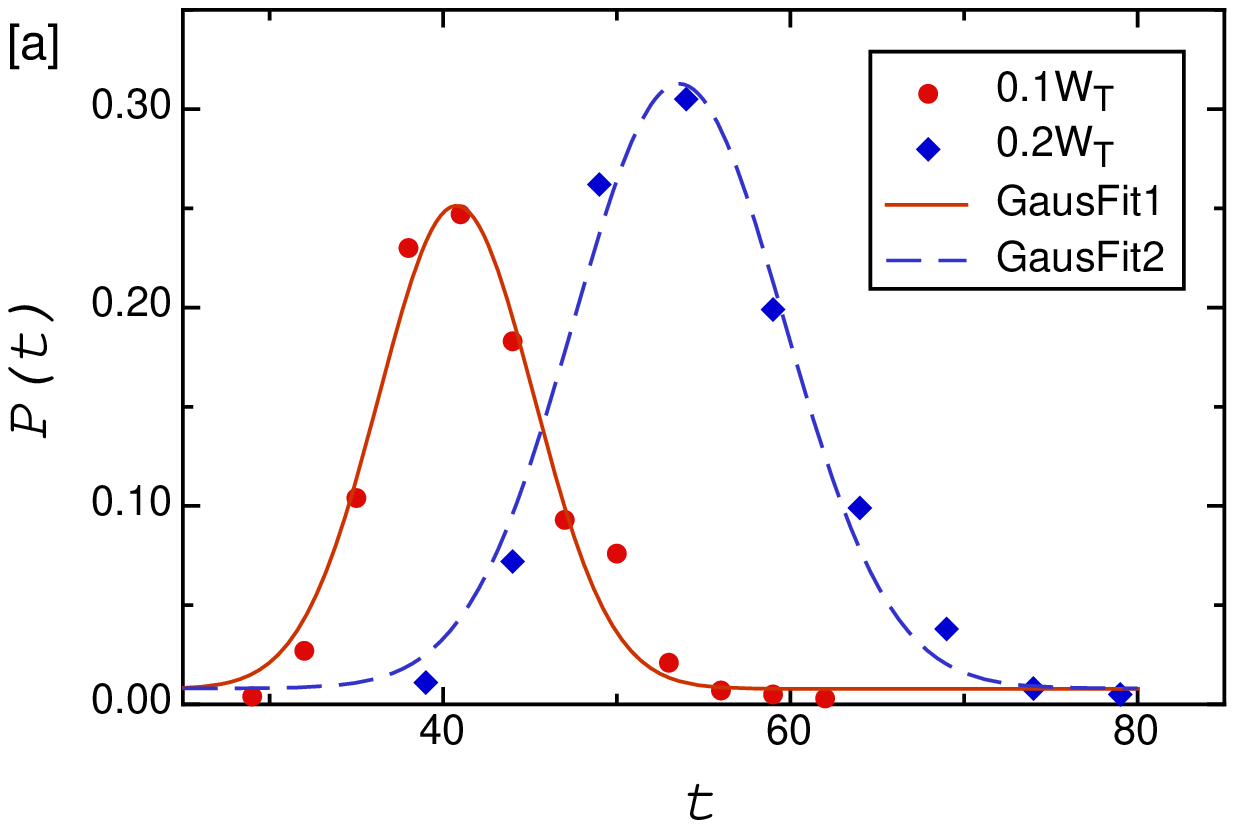}&
\includegraphics[width=0.45\columnwidth]{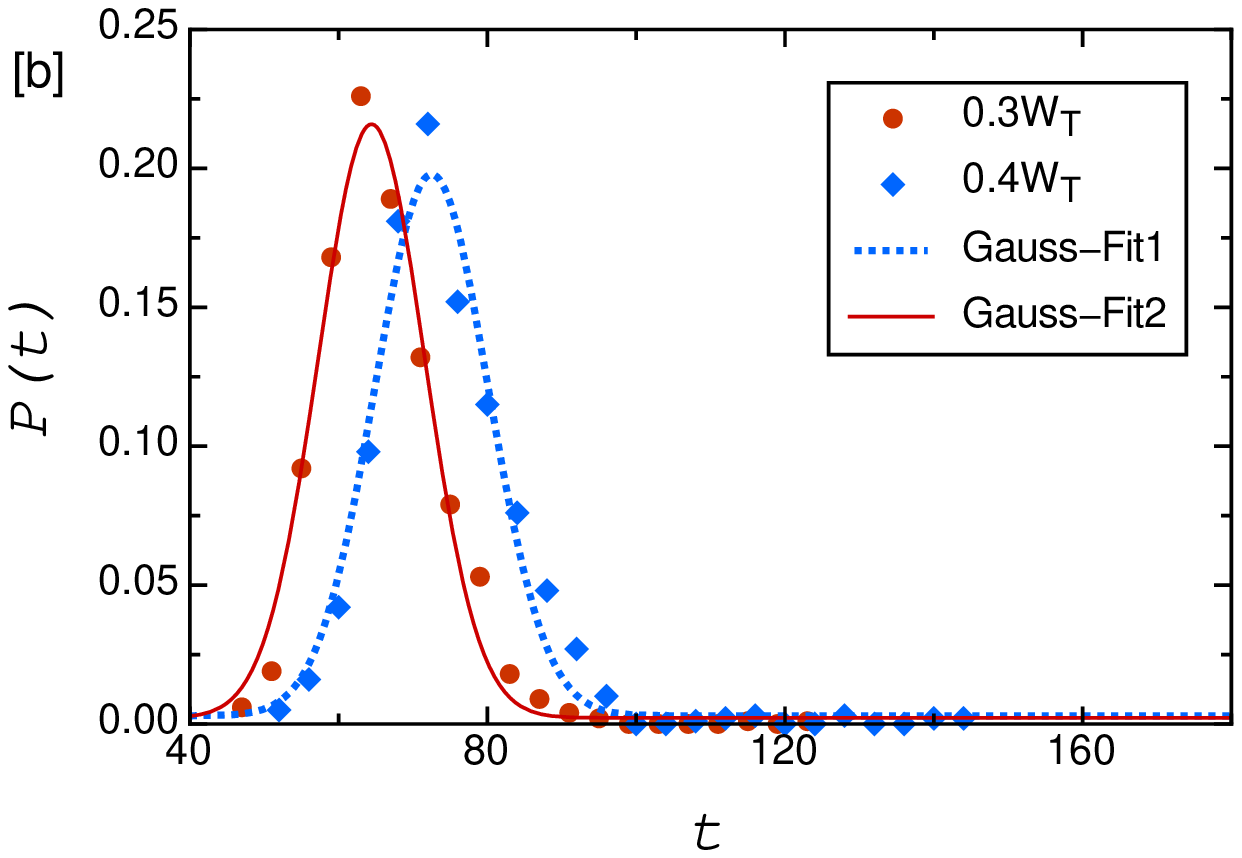}\\
\includegraphics[width=0.45\columnwidth]{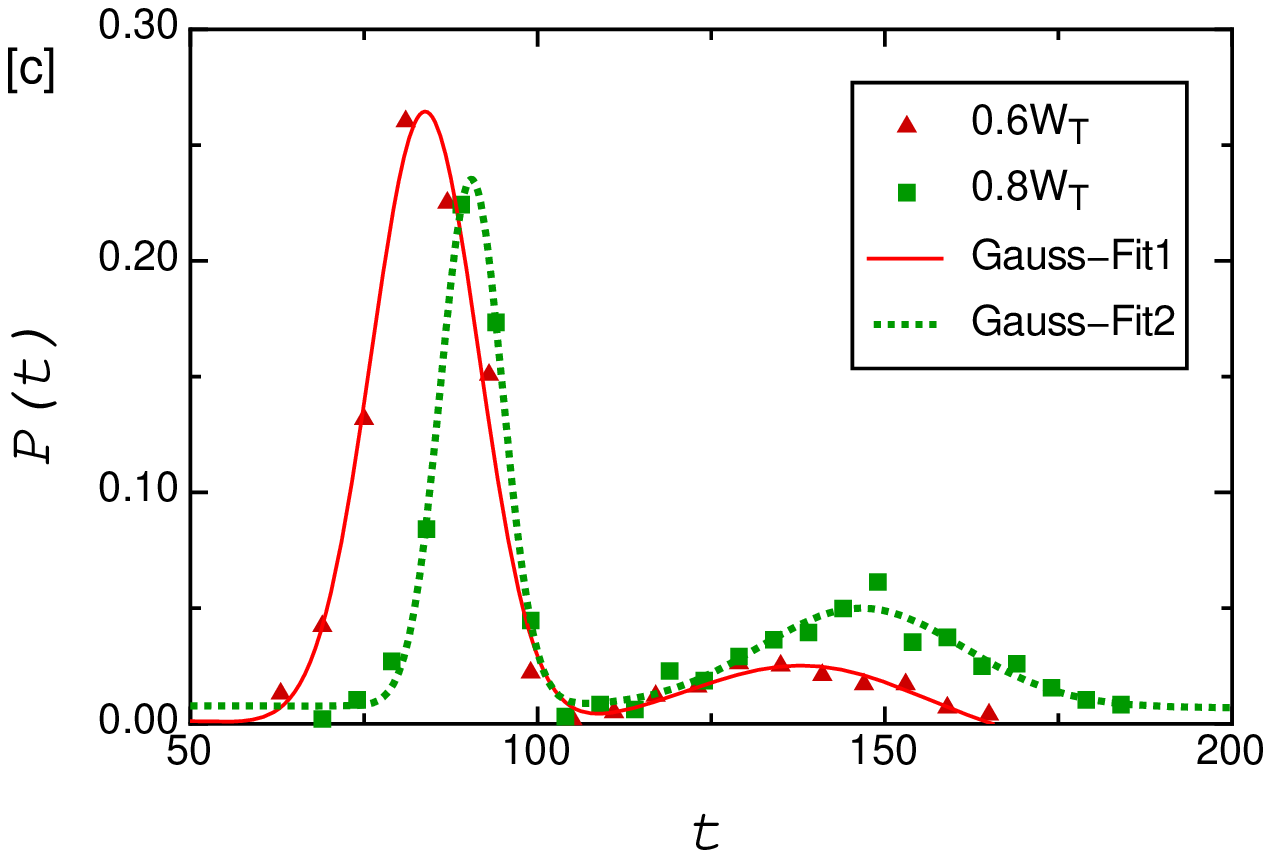}&
\includegraphics[width=0.45\columnwidth]{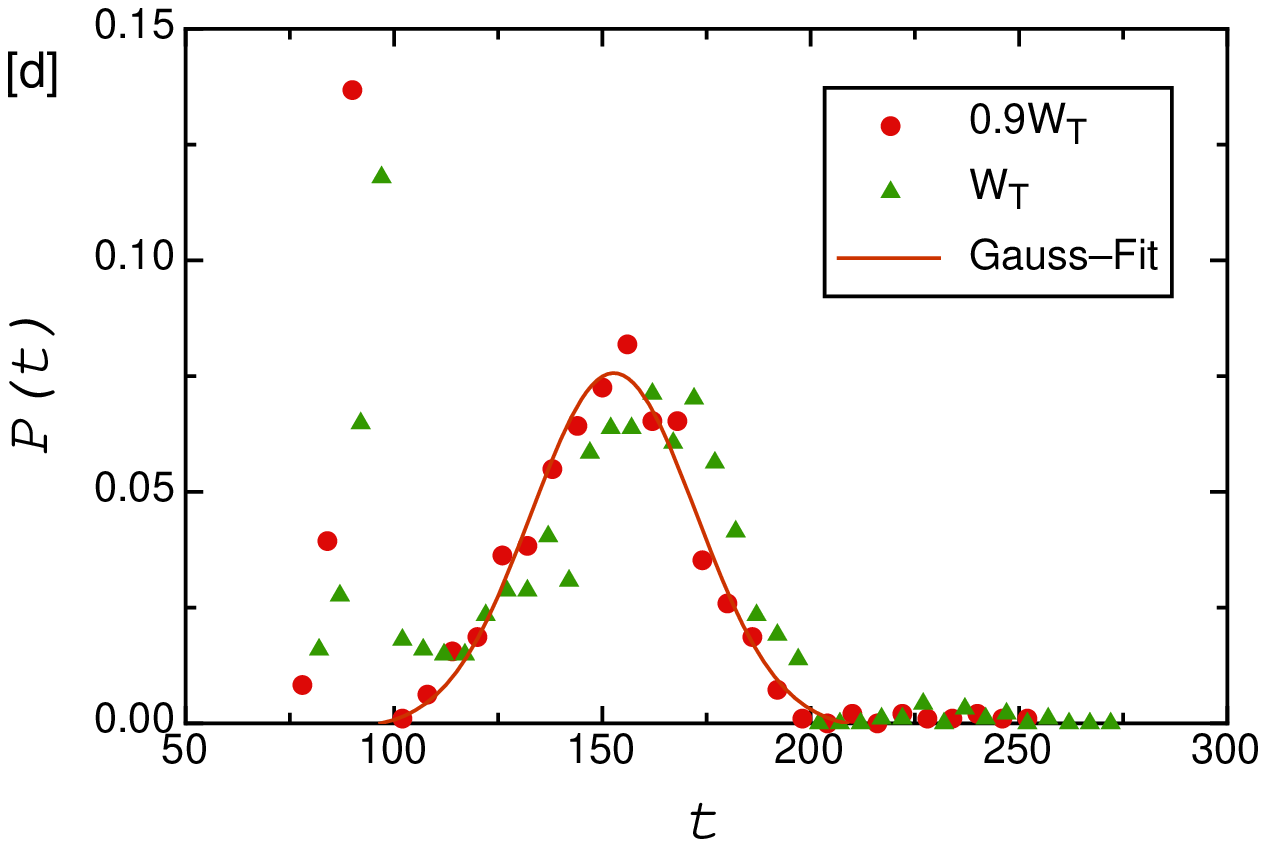}\\
\end{tabular}{}
\caption{Probability distributions of avalanche time $t$ for original lattice corresponding to $1000$ realizations of networks of side $M=100$, fit to Gaussian given by Eq.(2) when tested for weights equal to (a) $0.1W_T$ ($\sigma$=$4.46$, $\chi^2$=$0.0028$) and $0.2W_T$ ($\sigma$=$6.08$, $\chi^2$=$0.0024$), (b) $0.3W_T$ ($\sigma$=$6.92$, $\chi^2$=$0.0016$) and $0.4W_T$ ($\sigma$=$7.34$, $\chi^2$=$0.0024$), (c) $0.6W_T$ ($\sigma_1$=$7.75$, $\sigma_2$=$17.55$, $\chi^2$=$0.0013$) and $0.8W_T$ ($\sigma_1$=$4.46$, $\sigma_2$=$14.68$, $\chi^2$=$0.0062$), and (d) $0.9W_T$ ($\sigma$=$19.82$, $\chi^2$=$0.0035$) and $W_T$.}
\label{fig:origwt}
\end{center}
\end{figure}

The probability distribution of avalanche times $t$ for the original lattice is shown in Fig.\ref{fig:orig} for an ensemble of $1000$ successful weight transmissions for weight equal to trunk capacity $W_T$. In the original lattice no avalanche is seen when $t/M < 1$ as there are no paths with capacity greater than trunk capacity $W_T$ $i.$ $e.$ there are no successful transmissions for $t/M<1$. Also the avalanches of weight transmission in original lattice can cycle as much as thrice through the network. When these distributions for original networks of different sizes are scaled by their respective total number of layers $M$, they collapse on one another as shown in Fig. \ref{fig:orig}. Similar behavior is seen in Ref. \cite{janaki}.

The distribution of avalanche times $t$ for original lattice for test weights which are fractions of trunk capacity $W_T$, is different from that for weights equal to trunk capacity as shown in Fig. \ref{fig:origwt}. Fig. \ref{fig:origwt} shows the distribution $P(t)$ tested for weights ranging from to $0.1W_T$ to $0.9W_T$. In the distributions corresponding to $0.1W_T$ and $0.2W_T$ as shown in Fig. \ref{fig:origwt} (a), there is only one cycle in the form of a Gaussian peak, which can be expressed by the equation
\begin{equation}
P(t)=\frac{1}{{\sigma}{\sqrt {2\pi}}}\mathrm{exp}(-\frac{(t-a)^{2}}{2{\sigma}^{2}})
\end{equation}
where $a$ is a constant and $\sigma$ is the standard deviation. The values of $\sigma$ and chi-squared $\chi^2$ tested for the accuracy of the fits corresponding to $0.1W_T$ and $0.2W_T$ are shown in the caption of Fig. \ref{fig:origwt}. In the distributions corresponding to $0.3W_T$ and $0.4W_T$ the emergence of second cycle is seen as shown in Fig. \ref{fig:origwt} (b). This second cycle persists till $0.9W_T$, besides, it gradually attains the form of a new Gaussian peak as can be seen in Fig. \ref{fig:origwt} (c) and (d). In the distribution corresponding to $0.9W_T$, the first cycle completely loses its Gaussian form, and a new third cycle emerges out after the Gaussian peak corresponding to the second cycle, which is apparent from Fig. \ref{fig:origwt} (d).
Thus as the test weight start approaching the trunk capacity, we start seeing more and more transmissions which cycle more than once through the lattice.

\begin{figure}[ht]
\begin{center}
\begin{tabular}{cc}
\includegraphics[width=0.5\columnwidth]{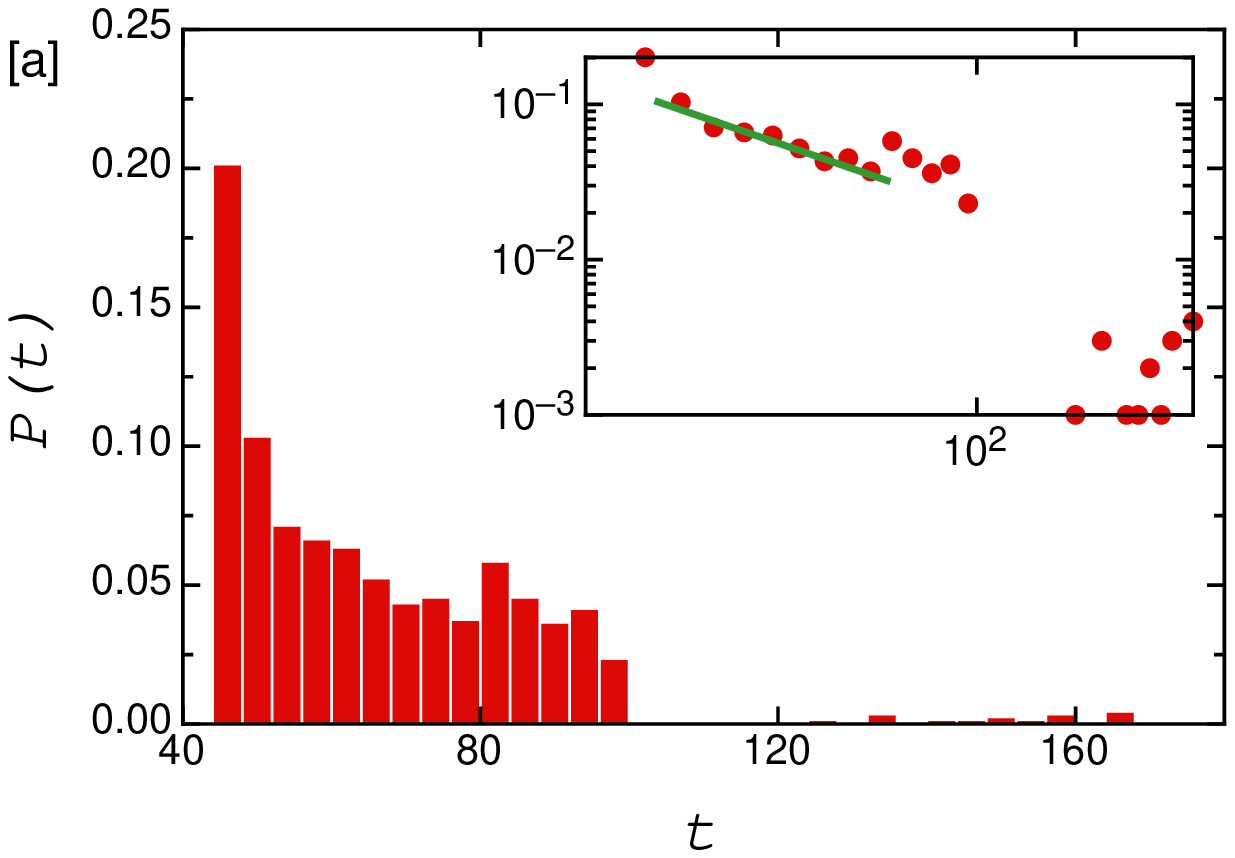}&
\includegraphics[width=0.5\columnwidth]{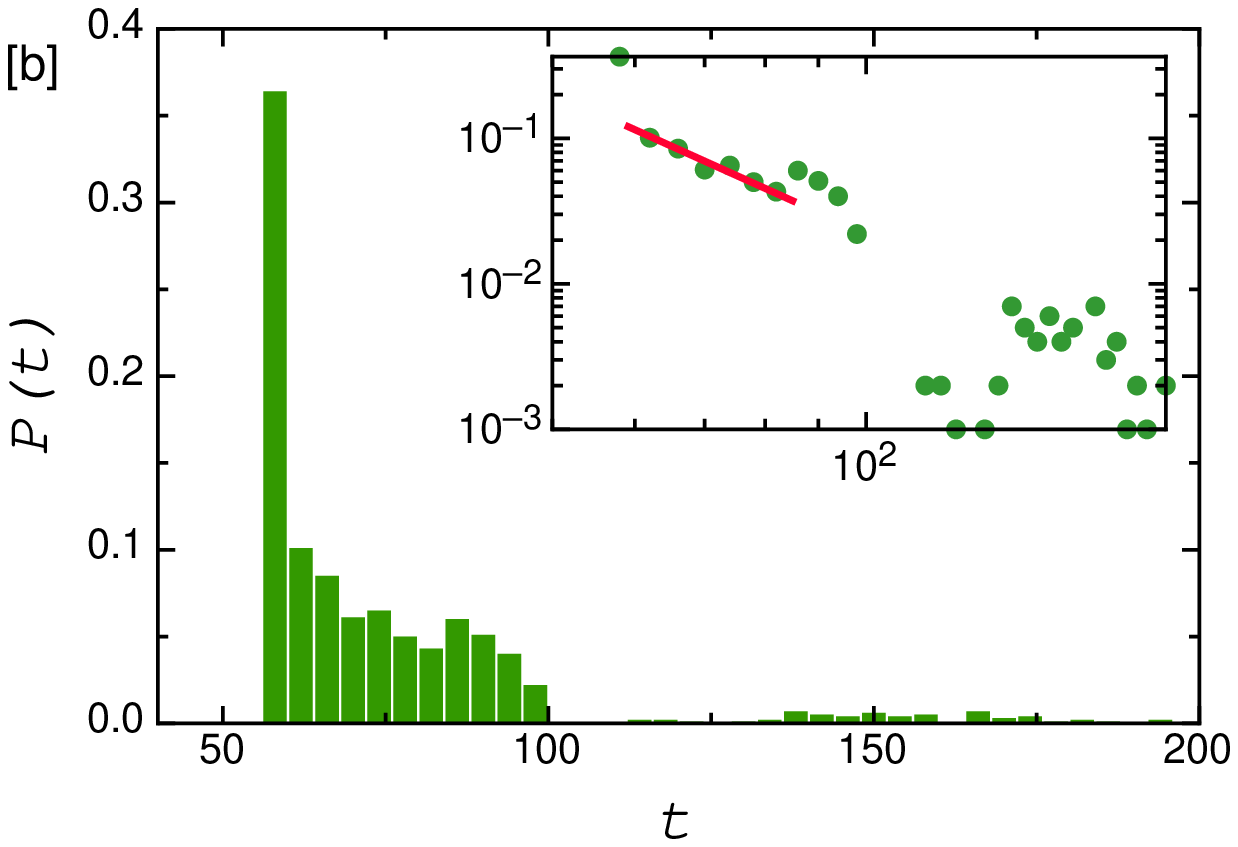}\\
\end{tabular}{}
\caption{Probability distributions of avalanche time $t$ corresponding to $1000$ realizations for the $V-$ lattice network of side $M=100$ when tested for weights equal to (a) $0.1W_T$, and (b) $0.2W_T$. Small regimes for $0.1W_T$ (as shown in inset of (a)) and $0.2W_T$ (as shown in inset of (b)) display power law behavior with exponent $\alpha=2.45$ and $\chi^2=21.136$, and $\alpha=2.96$ and $\chi^2=52.197$ respectively.}
\label{fig:vlat}
\end{center}
\end{figure}

The distribution of avalanche times for the $V-$ lattice is found to be quite different from that of the original lattice. It does not show any systematic behavior for weights equal to trunk capacity $W_T$ $i.$ $e.$ distinct lattice realizations show distinct behavior. However, the $V-$ lattice does show interesting behavior for avalanche time distributions when it is tested for weights less than its trunk capacity. Fig. \ref{fig:vlat} displays the probability distribution of avalanche times for the $V-$ lattice ($1000$ realizations). It is clear from the figure that the $V-$ lattice displays a power law behavior of the form $P(t)\sim t^{-\alpha}$ with exponent $\alpha=2.45$ and $\alpha=2.96$ when weights equal to $0.1W_T$ (Fig. \ref{fig:vlat} (a)) and $0.2W_T$ (Fig. \ref{fig:vlat} (b)) are placed in the first layer, respectively. The power law regime in the distribution of avalanche times  gradually starts disappearing when the distribution is tested with values of weights placed approaching trunk capacity, and at the trunk capacity no stable behavior in the distribution for different realizations is seen. The difference in the behavior of the distribution of the $V-$ lattice from the original lattice is due to the presence of a unique and asymmetric cluster, called the $V-$ cluster, in the $V-$ lattice network. The $V-$ cluster includes sites of all levels of capacity. Hence transmissions on the lattice can achieve success at any one of the layers. This accounts for the power law distribution and thus brings criticality to the distribution of avalanche times $t$. When the distribution for the $V-$ lattice is tested for weights higher than $0.2W_T$ the power law regime of the avalanche time distribution starts dying down slowly much before the instability in distribution is seen at trunk capacity $W_T$. This behavior of the existence and subsequent disappearance of the power law regime in the distribution is one of the indications that the $V-$ lattice is indeed a critical case of the original lattice. Such behavior has not been seen in any of realization of the original lattice except the $V-$ lattice. Even two connectivity strategies studied by Janaki $et$ $al.$ \cite{janaki} does not show the existence of such behavior.

\begin{figure}[ht]
\begin{center}
\includegraphics[width=0.6\columnwidth]{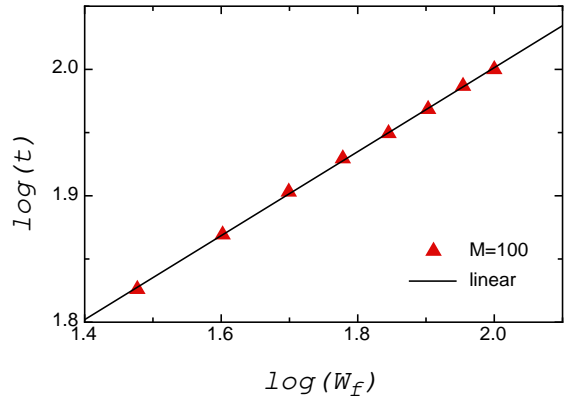}
\caption{Avalanche time $t$ displays power law behavior $t\sim {W_{f}}^{\beta}$ with $\beta=0.33$ and $\chi^2=0.00001$ against weights placed as percentage increase in trunk capacity for $V-$ lattice for $M=100$.}
\label{fig:vlat_time}
\end{center}
\end{figure}

The study of avalanche times $t$ of successful weight transmissions along the trunk path in the $V-$ lattice is shown in Fig. \ref{fig:vlat_time}. From this figure it is clear that the avalanche time $t$ along the trunk is also governed by a power law $t\sim W_f^{\beta}$ with $\beta=0.33$, against $W_f$ which are test weights ranging from $0.1W_T$ to $W_T$.

\section{Conclusions}

Threshold phenomena propagating on branching hierarchical lattices are of interest
in a diverse variety of application contexts. Models of this kind have
been proposed in the context of respiration networks \cite{inflation, inflation2}, voter models \cite{voter}, 
granular media \cite{copper}, power networks \cite{dobson}, river networks \cite{river}, as well as directed percolation \cite{domany} contexts. Avalanche distributions in these contexts have significance
for ventilation strategies for respiratory networks, opinions or
preferences cascading on voter networks as well as numerous percolation
contexts. It is interesting to note the significance of the $V-$ lattice
configuration as the critical configuration in these contexts. It is
clear from our results that a $V-$ shaped structure which spans the breadth of the
lattice at one extreme, with the trunk, i.e. the collection of the
strongest sites along one arm, is best able to support cascades of all
possible scales. The utility of this configuration for optimizing the
connectivity structure of networks deserves to be explored further. We
hope to explore some of these directions in future work.

\acknowledgments
AD thanks the University Grants Commission, India, for a fellowship.

\bibliographystyle{pramana}
\bibliography{references}

\end{document}